\documentclass[aps,prl,twocolumn,longbibliography,superscriptaddress,floatfix,10pt]{revtex4-2}
\usepackage[english]{babel}
\usepackage{amsmath,amssymb,bbm,mathrsfs,bm,braket,color,graphicx,comment,amsfonts,dsfont}
\usepackage[colorlinks,linkcolor=blue,citecolor=blue,urlcolor=blue]{hyperref}
\usepackage[mathscr]{euscript}
\usepackage{physics}
\usepackage{xcolor}
\usepackage[normalem]{ulem}
\usepackage{bm}
\usepackage{orcidlink}
\usepackage{multirow}
\usepackage{microtype}
\usepackage{bbold}
\usepackage{pifont}

\newcommand{\bs}[1]{\boldsymbol{#1}}

\begin{document}

\title{Quantum geometry, localization, and topological bounds of spin fluctuations}
\author{Carlos Saji}
\affiliation{Departamento de F\'isica, Universidad de Santiago de Chile, Santiago, Chile.}
\author{Roberto E. Troncoso}
\affiliation{Departamento de Física, Facultad de Ciencias, Universidad de Tarapacá, Casilla 7-D, Arica, Chile}

\begin{abstract}
We study how topological crystalline defects—dislocations—reshape the real-space quantum geometric tensor and act as tunable sources of quantum geometry. We show that dislocations strongly enhance the quantum metric, establishing a direct link between lattice topology and the Hilbert-space geometry of states. We characterize the quantum geometry of topological magnons in ordered arrays of dislocations, demonstrating that defect-induced geometric enhancement controls their localization and topological protection. In disordered arrays, dislocation-driven geometry expands the accessible topological phase space and enables transitions to disorder-induced topological phases. Our results identify the quantum metric as a tunable bridge between crystalline topology, magnonic excitations, and emergent topological matter in aperiodic solid-state and synthetic systems.
\end{abstract}

\maketitle

\textit{Introduction--} Quantum geometry—the geometric structure of Hilbert space states—is central to understanding fundamental phenomena in condensed matter physics \cite{Provost1980,Resta2011}. In parameter space, quantum geometry is fully characterized by the quantum geometric tensor (QGT), also known as the Fubini–Study metric. The imaginary part of the QGT corresponds to the well-studied Berry curvature, which plays a central role in understanding adiabatic phase evolution and transport phenomena such as the anomalous Hall effect \cite{NagaosaRMP2010}, and serves as a cornerstone in the study of topological matter \cite{NagaosaRMP2010,XiaoRMP2010,HasanRMP2010,QiRMP2011}. In contrast, much less attention has been given to the real component of the QGT—the quantum metric—which quantifies the local distance between nearby quantum states in the Hilbert space \cite{Provost1980,Resta2011}. Quantum metric is crucial in diverse material properties and phenomena, bridging geometric and topological aspects of quantum matter \cite{TomaPRL2023,Liu2024,Yu2025}. It contributes to intrinsic nonlinear conductivities in noncentrosymmetric systems \cite{AhnPRX2020,WatanabePRX2021,Gao2023,Wang2023,WangPRL2023,DasPRB2023,KaplanPRL2024,Jiang2025,ulrich2025}, determines the stability of fractional Chern insulators \cite{RoyPRB2014,WangPRL2021,EstiennePRR2023}, and governs flat-band superconductivity through geometric pairing mechanisms \cite{TormaPRB2018,XiePRL2020,JulkuPRB2020,AbouelkomsanPRR2023}. Moreover, it contributes to orbital magnetism \cite{Slot2023,zhang2025,to2025} and provides a quantitative measure of the localization and spatial extent of Wannier electronic states \cite{MarzariPRB1997}.

The Berry curvature underpins transport properties and the formation of topological states in classical magnetism \cite{McClarty2022,Zhuo2023,Onose2010}. Magnetic order supports magnonic excitations—the bosonic counterparts of spin waves—that can host topological states in the presence of magnetic textures \cite{Schutte2014,McClarty2022} or spin–orbit coupling \cite{Zhuo2023}. In this regime, the corresponding Bogoliubov–de Gennes Hamiltonian exhibits magnon bands with nonzero Chern numbers, i.e., momentum-space integrals of the Berry curvature, leading to chiral edge states \cite{McClarty2022,Zhuo2023}. Beyond this ideal framework, when magnons interact with crystalline defects such as disorder (magnetic or structural) or lattice dislocations, new tools are required to characterize the real-space topology \cite{Wang2020}. Crystal dislocations generate local strain fields that break translational symmetry, modifying the magnon band structure and producing defect-bound states that are typically localized around dislocations \cite{Saji2025}, depending on their spatial distribution and interactions. When parity symmetry is broken, magnonic states bound to the dislocation can become $\mathbb{Z}_2$ topologically protected \cite{Saji2025}. Such topological protection is not exclusive to bosonic excitations, as analogous phenomena have been observed in electronic systems \cite{Ran2009}. While we focus on magnons, the dislocation-induced enhancement of the quantum metric is a universal geometric response to lattice defects, not restricted to bosonic excitations and applicable, in principle, to fermionic topological systems.

In this Letter, we demonstrate that crystalline dislocations lead to a pronounced enhancement of the quantum metric, revealing a fundamental link between lattice topology and quantum geometry. The presence of dislocations reshapes the local Hilbert-space geometry, amplifying geometric responses associated with the underlying magnonic band structure. {This enhancement, present in both ordered and disordered arrays of dislocations, correlates with an increased spatial extent of the magnonic states, consistent with the established relation between the quantum metric and Wannier-state localization.} Crystalline defects thus act as geometric degrees of freedom that tune the quantum metric, control magnon localization and topological phases.

\textit{Spin fluctuations.--} Consider a ferromagnetic spin system localized on a two-dimensional (2D) hexagonal lattice. The low-energy spin fluctuations (magnons) about the ground state are described by the real-space and non-interacting Bogoliubov-de Gennes (BdG) Hamiltonian,
\begin{align}\label{eq: BdG-Hamiltonian}
\hat{\cal H}=\sum_{{\bs r}{\bs r}'}\hat{\Psi}^{\dagger}_{\bs r} {\cal M}_{{\bs r}{\bs r}'}\hat{\Psi}_{\bs r'},
\end{align}
where ${\cal M}$ is a hermitian $2N\times 2N$-matrix. The tight-binding BdG Hamiltonian originates from bosonization of a total spin Hamiltonian \cite{WangJPD2018,Wang2020,Saji2025}, $\hat{\cal H}_S=\hat{\cal H}_E+\hat{\cal H}_{D}+\hat{\cal H}_{pd}+\hat{\cal H}_A$, containing exchange interaction ($\hat{\cal H}_E$), Dzyaloshinskii-Moriya coupling ($\hat{\cal H}_D$), pseudo-dipolar interaction ($\hat{\cal H}_{pd}$), and easy-axis anisotropy ($\hat{\cal H}_A$). The Nambu spinor ${\hat{\bs\Psi}}=({\hat {\bs a}},\hat{\bs a}^{\dagger})^T$ of bosonic annihilation ($\hat a_{\bs r}$) and creation ($\hat a^{\dagger}_{\bs r}$) operators, at site ${\bs r}$, represent magnonic excitations, which are introduced via Holstein-Primakoff (HP) bosons \cite{HolsteinPR1940}. Around the {$z$-axis oriented} FM state, the spin-to-magnon mapping reads ${S}^{+}_{\bs r}=\left(2S-a^{\dagger}_{\bs r}a_{\bs r}\right)^{1/2}a_{\bs r}$, ${S}^{-}_{\bs r}=\left(2S-a^{\dagger}_{\bs r}a_{\bs r}\right)^{1/2}a^{\dagger}_{\bs r}$, and ${S}^{z}_{\bs r}=S-a^{\dagger}_{\bs r}a_{\bs r}$, where the Hamiltonian \ref{eq: BdG-Hamiltonian} is obtained expanding the spin operators as a series in $1/S$. The bosonic Hamiltonian is para-diagonalized by the Bogoliubov transformation $(a_{\bs r},\dots,a^{\dagger}_{\bs r})^T={\text T}_{\bs r\bs r'}(\alpha_{\bs r'},\dots,\alpha^{\dagger}_{\bs r'})^T$, by the Colpa algorithm \cite{colpa1978diagonalization}, with ${\text T}$ the paraunitary transformation that satisfy ${\text T}^{\dagger}\Sigma_z{\text T}=\Sigma_z$ to guarantee the bosonic commutation relation $\left[{\bs \alpha},{\bs \alpha}^{\dagger}\right]=\mathbb{1}\otimes\sigma_z=\Sigma_z$ (with $\sigma_z$ the Pauli matrix). Therefore, the diagonalized magnon Hamiltonian is written in momentum space as $\hat{\cal H}=\sum_{n,{\bs k}}{\cal E}_{{\bs k}n}\alpha^{\dagger}_{{\bs k}n}\alpha_{{\bs k}n}$, with ${\cal E}_{{\bs k}n}$ the energy for the $n^{\text{th}}$-band.

\textit{Magnonic quantum geometry.--} In translationally invariant systems, the QGT is defined with respect to crystal momentum, by $\mathcal{Q}^{nm}_{\mu \nu}({\bs k})=\langle\partial_{\mu} u^{n}_{\bs k}|\mathbb{1}-P_{\bs k}|\partial_{\nu} u^{m}_{\bs k}\rangle$, where $\mu,\nu$ are spatial direction indices. The $n$th Bloch band $u^{n}_{\bs k}$, where $n=1, \dots, N$, form a set of $N$ orthonormal row vectors of the Hilbert space, the momentum derivatives are denoted by $\partial_{\mu} = \partial/\partial {\bs k}_{\mu}$, with $\mu,\nu\in\{x,y\}$, and $P({\bs k})=\sum_{n\in\text{occ}}\left|u_{n\bs k}\right\rangle\left\langle u_{n\bs k}\right|$ is the projector onto occupied bands. Tracing over all
occupied bands the QGT decomposes as $\mathcal{Q}_{\mu \nu}({\bs k})=\text{Tr}\left[\mathcal{Q}^{nm}_{\mu \nu}({\bs k})\right]= g_{\mu \nu}({\bs k})+i \Omega_{\mu \nu}({\bs k})$, where $g_{\mu \nu}({\bs k})= \text{Tr} \left[\partial_{\mu}P\partial_{\nu}P \right]/2$ is the quantum metric (real and symmetric part) and $
\Omega_{\mu \nu}({\bs k})= i\,\text{Tr} \left[P[\partial_{\mu}P,\partial_{\nu}P ]\right]$ the Berry curvature (imaginary and antisymmetric part). The tensor $\mathcal{Q}_{\mu \nu}$ is positive semidefinite, which is related to the non-negative norm of eigenstates. Importantly, $\text{Tr}\left[g_{\mu \nu}({\bs k})\right]$ represents the localization functional for maximally localized Wannier functions, which highlights the connection between quantum geometry and localization \cite{MarzariPRB1997}. Integrating over the Brillouin zone (BZ) the Berry curvature defines the Chern number $\mathcal{C}= {\pi}^{-1}\int_{BZ} \Omega_{xy} ({\bs k})\ d^{2}{\bs k}$ which, under certain circumstances is quantized, in contrast to the integrated quantum metric (IQM) $\mathcal{G}= {(2\pi)}^{-1}\int_{BZ} \text{Tr}\left[g_{\mu \nu}({\bs k})\right]\ d^{2}{\bs k}$ or the quantum volume $\mathcal{V}_{g}= {\pi}^{-1}\int_{BZ} \sqrt{\text{det}\left[g_{\mu \nu}({\bs k})\right]}\ d^{2}{\bs k}$. However, these quantities hold the global inequality $\mathcal{C} \leq \mathcal{V}_{g}\leq \mathcal{G}$, that establishes a topological bound to the quantum geometry of states. In turn, this enables one to establish a lower bound on the geometric contribution to quantities such as the superfluid weight \cite{PeottaNC2015}, the static structure factor \cite{OnishiPRL2024}, and optical gaps \cite{OnishiPRX2024}, among others.

\begin{figure}[!htb]
\centering
\includegraphics[width=\linewidth]{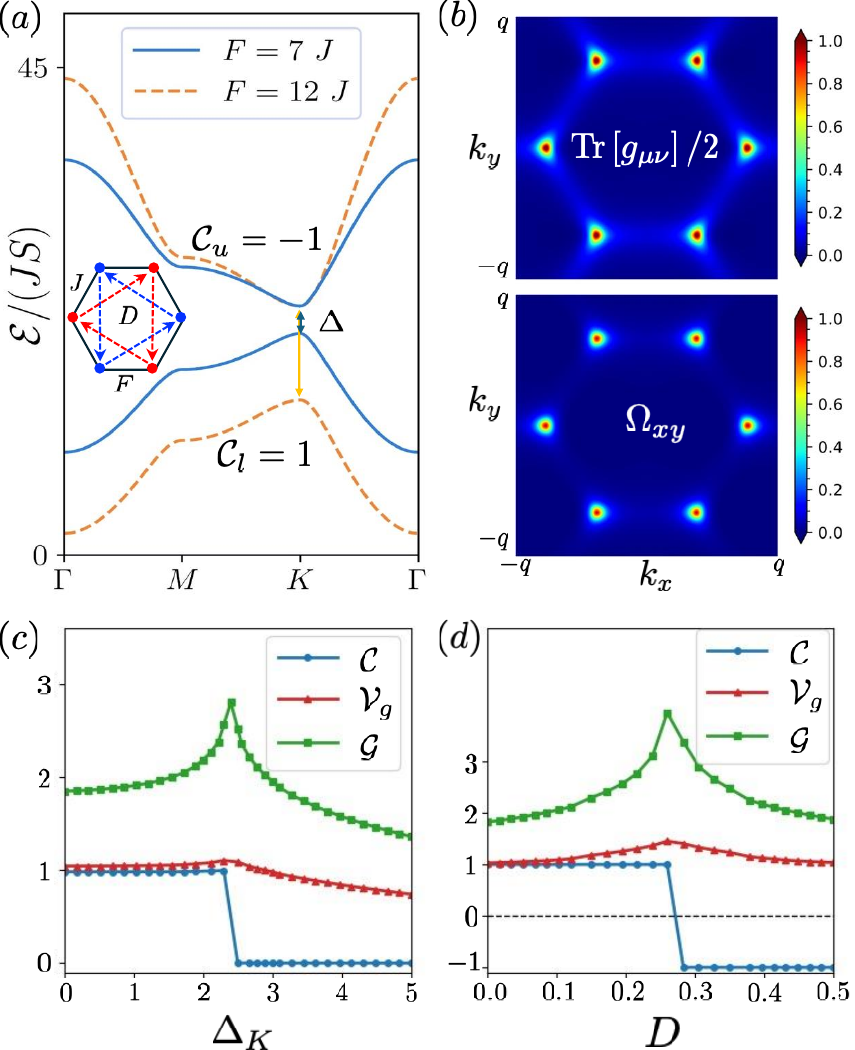}
\caption{(a) Magnonic band spectrum with $D=\Delta_{K}=0$ with the topological gap. At the inset scheme of magnetic couplings acting between nearest ($J$ and $F$) and next-nearest ($D$) neighbours indicated by the oriented colored bonds. (b) QM density in momentum space $\mathrm{Tr}[g_{\mu\nu}(\boldsymbol{k})]$ and Berry curvature $\Omega(\boldsymbol{k})$ for the case (a). Quantum metric $\mathcal{G}$, quantum volume $\mathcal{V}_{g}$ and Chern number $\mathcal{C}$ as a function of $\Delta_{K}$ (c) and DMI (d) for the lower band.  In all plots, the parameters are set to $J=1$, $K=20J$ and $F=7J$.}
\label{Figure1}
\end{figure}
\begin{figure*}
\centering
\includegraphics[width=\linewidth]{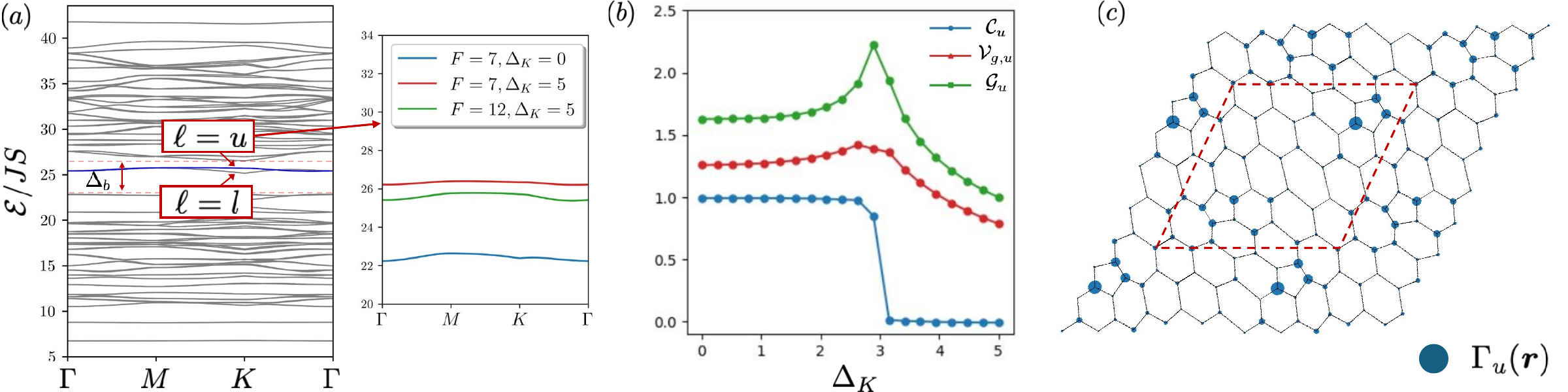}
\caption{(a) Bulk magnon structure of the magnetic triangular arrays of dislocations, along first BZ loop $\Gamma$-$M$-$K$-$\Gamma$, for the parameters $J=1$, $K=20J$, $F=7J$ and $\Delta_K=5J$. Inside the gap $\Delta_b$ lies a pair of magnonic states. The higher-energy mode of this pair is highlighted in blue, and zoomed-in for other parameter values. (b) Trace of the quantum metric $\mathcal{G}_u$ (green), quantum volume $\mathcal{V}_{g,u}$ (red) and Chern number $\mathcal{C}_u$ (blue), as a function of $\Delta_{K}$ { for the dislocation band $\ell=u$ highlighted in blue in panel (a).} (c) Real-space distribution of the magnonic mode $\ell=u$, highlighted in blue at (a), represented by $\Gamma_u(\bs r)$. 
}
\label{Fig2:crystal-dislocations}
\end{figure*}
In the absence of topological defects, the lattice realizes the full point-group symmetry $D_6$, and the magnon Hamiltonian reduces to the well-studied two-band model, $\hat{\cal H}=\sum_{{\bs k}}\hat{\Psi}^{\dagger}_{\bs k} {\cal M}_{{\bs k}}\hat{\Psi}_{\bs k}$, which exhibits nontrivial bulk topology \cite{WangJPD2018,Wang2020}. Here, 
\begin{equation}
{\cal M}_{\bs k}=\left(\begin{matrix}
M_{\cal A} & N_{\bs k} \\
N^{\dagger}_{\bs k} & M_{\cal B}\\
\end{matrix}\right), \qquad
N_{\bs k}=\left(\begin{matrix}
f_{\bs k} & g_{+,\bs k} \\
 g_{-,\bs k} & f_{\bs k} \\
\end{matrix}\right),
\end{equation}
where $M_{\cal A}=M^{\xi}_{\cal A}\delta_{\xi\xi}$ and $M_{\cal B}=M^{-\xi}_{\cal B}\delta_{\xi\xi}$ being $\xi=\pm$, with $M^{\pm}_{\zeta}=SK_{\zeta}+3JS\pm d_{\bs k}$, $K_{\zeta}=K\pm\Delta_K$ and $\zeta\in {\cal A,B}$, respectively, $d_{\bs k}=2DS\sum_j\sin ({\bs k}\cdot {\bs b}_j)$, $f_{\bs k}=-S\left(J+{F}/{2}\right)\sum_j e^{i{\bs k}\cdot {\bs a}_j}$, $g_{\pm,\bs k}=F S\sum_j e^{\pm2i\theta_j}e^{i{\bs k}\cdot {\bs a}_j}/{2}$, and ${\bs a}_j({\bs b}_j)$ the position of nearest (next-nearest) neighbors. The bulk magnon spectrum, shown in Fig. \ref{Figure1}(a), exhibits a gap at the $\Gamma$ point—signaling the absence of Goldstone modes—and at the Dirac points, where the gap is of topological origin. The two magnon bands carry opposite Chern numbers, and the resulting topological gap $\Delta$ is governed by a combined effect of the Dzyaloshinskii–Moriya interaction, pseudo-dipolar coupling, and anisotropy strength \cite{WangJPD2018,Wang2020}. The QGT is determined from
\begin{align}\label{eq:QGT}
\mathcal{Q}_{\mu \nu}^{(n)}({\bs k})= \sum_{{m\neq n}} \frac{\left\langle u_{n\bs k}\right| \partial_{k_{\mu}} \mathcal{H}\left|u_{m\bs k}\right\rangle\left\langle u_{m\bs k}\right| \partial_{k_{\nu}} \mathcal{H}\left|u_{n\bs k}\right\rangle}{\sigma_{n}\sigma_{m}\left(\sigma_{n}{\cal E}_{n\bs k}-\sigma_{m}{\cal E}_{m\bs k}\right)^2},
\end{align}
where $\sigma_{n}=(\zeta)_{nn}$. {Eq.~(\ref{eq:QGT}) defines the quantum geometric tensor of a single
magnon band $n$. The index $m$ runs over all remaining bands, $m\neq n$, and accounts for virtual interband contributions through
the Feynman--Hellmann representation.} The distribution in the BZ of quantum metric, $\mathrm{Tr}\left[g_{\mu\nu}(\bs k)\right]$, and Berry curvature, $\Omega_{xy}(\bs k)$, are shown in Fig. \ref{Figure1}(b). The interplay between topology and quantum geometry is summarized in Figs. \ref{Figure1}(c) and (d), which plot the Chern number (blue), quantum volume (red), and the IQM (green), versus tuning parameters. The Chern number remains quantized in the topologically nontrivial regime, and exhibits a sharp jump—to ${\cal C}=0$ for increasing $\Delta_K$ and to ${\cal C}=-1$ for increasing $D$. Near the topological phase transition, $V_g$ develops a mild cusp, while ${\cal G}$ shows a pronounced peak, indicative of enhanced quantum geometry and magnon-wavefunction delocalization as the topological gap closes \cite{MarzariPRB1997}.

\textit{Crystal of dislocations.--} We now consider the presence of an array of dislocations in the hexagonal lattice. These defects form a triangular pattern, see Fig. \ref{Fig2:crystal-dislocations}, that locally distorts the crystal environment, consistent with a threefold ($C_3$) rotational symmetry. The magnetic unit cell of the classical ground-state comprises more than one spin, thus the lattice site index ${\bs r}$ is partitioned as ${\bs r}={\bs R}+{\bs r}_j$. Here ${\bs R}$ is a Bravais lattice vector and the ${\bs r}_j$’s label the spins within the magnetic unit cell. Under this relabeling, the HP bosons are now written as ${a}_{\bs kj}$ and are related to their lattice Fourier transforms through $a_{\bs Rj}=(1/\sqrt{N})\sum_{\bs k}e^{i{\bs k}\cdot({\bs R}+{\bs r}_j)}a_{\bs kj}$, with $N$ being the total number of magnetic unit cells. Upon Fourier transform, the magnon Hamiltonian in the ordered array of dislocations is then given by $\hat{\cal H}_{od}=\sum_{{\bs k},ij}\hat{\Psi}^{\dagger}_{\bs ki} {\cal M}_{ij}({\bs k})\hat{\Psi}_{\bs kj}$, where $\hat{\Psi}_{\bs kj}=(a_{\bs kj},a^{\dagger}_{\bs kj})^T$. Following the standard procedure, we diagonalize the bosonic Hamiltonian by a linear Bogoliubov transformation.

The bulk magnon spectrum is shown in Fig.~\ref{Fig2:crystal-dislocations}(a) for $\Delta_K=5J$. The two lowest-energy bands, below ${\cal E}=10JS$, are remarkably flat, reflecting non-propagating modes with vanishing group velocity. Such flatness suppresses kinetic energy, thereby enhancing interaction effects and favoring the emergence of correlated phases \cite{PeottaNC2015, TormaPRB2018}. We also identify a bulk gap $\Delta_b$ between ${\cal E}=23JS$ and $26JS$, within which a pair of magnonic states (labeled as $\ell=l, u$) becomes gapped once parity symmetry is broken. Note that the main difference is located near the $K$ point. {The higher-energy state of this pair ($\ell=u$) is highlighted in blue in Fig. \ref{Fig2:crystal-dislocations}(a) and zoomed-in for other parameter values.}
The spatial distribution of these states is determined from the overlap $\Gamma_{\ell}({\bs r})=\left|\langle GS|a_{\bs r }\alpha^{\dagger}_{\ell}|GS\rangle\right|^2$, between the local excitation and the eigenstates, with $|{GS}\rangle$ the ground-state of the magnon Hamiltonian. {Figure \ref{Fig2:crystal-dislocations}(c) shows the real-space profile of the mode $\ell=u$, denoted by $\Gamma_u(\bs r)$.
Although this eigenmode is spatially concentrated around the dislocation cores \cite{Saji2025}, the periodic arrangement of dislocations gives rise to an extended Bloch band. Consequently, the band ($\ell=u$) can be characterized by a well-defined Chern number.
The nonzero Chern number in Fig. \ref{Fig2:crystal-dislocations}(b) originates from the momentum-space quantum geometry of the corresponding Bloch band.}
\begin{figure*}[!htb]
\centering
\includegraphics[width=\linewidth]{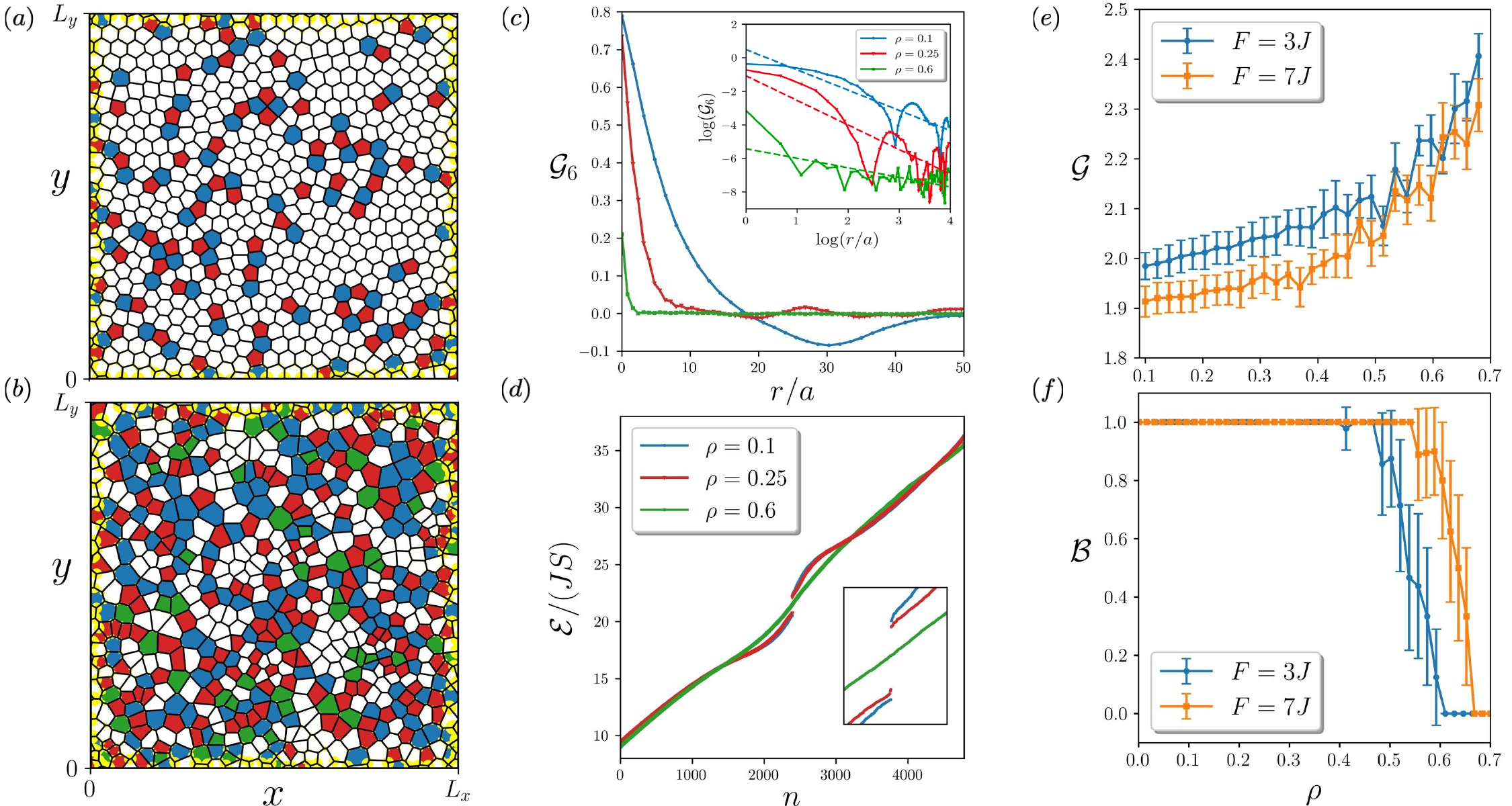}
\caption{Geometric and topological properties of magnonic states in the presence of randomly distributed dislocations on a hexagonal lattice. Examples of systems with low ($\rho=0.25$) and high ($\rho=0.6$) dislocation concentrations are shown in panels (a) and (b), respectively, for $L_x=L_y=40 a$ and $a$ the lattice constant of the perfect crystal, {where pentagons and heptagons are colored in red and blue, respectively.} Panel (c), {together with the inset shown on a logarithmic scale}, characterizes the corresponding degree of crystalline disorder via the bond-orientational correlation function ${\cal G}_6(r)$ for different dislocation densities $\rho=0.1$, $0.25$, and $0.6$. At high concentrations, it exhibits a power-law decay, ${\cal G}_6(r)\sim r^{-\eta}$, with $\eta=2.5$ and $\eta=3.1$ for $\rho=0.25$ and $\rho=0.6$, respectively, signaling the emergence of a Hexatic phase. {(d) Magnonic band spectrum determined for various dislocation concentrations. } { The quantum metric ${\cal G}$ and the Bott index ${\cal B}$ associated with the lower spectral sector are plotted in panels (e) and (f), respectively, as functions of the dislocation concentration for $F/J=3$ and $5$.} The local quantum metric $g(r)$, Eq. (\ref{eq:G_real_space}), computed for $\rho=0.25$ and $\rho=0.6$, is displayed (in yellow) at panels (a) and (b), respectively. The transition to a topologically trivial phase (${\cal B}=0$) coincides with a pronounced enhancement of the quantum metric, indicating the formation of {highly delocalized magnonic states}.}
\label{Fig3: Gas-dislocations}
\end{figure*}

The ordered array of dislocations gives rise to strain-induced modifications of the band geometry and quantum metric tensor, shown at Fig. \ref{Fig2:crystal-dislocations}(b), signaling increased Bloch-state overlap and enhanced delocalization of magnonic states. 
{The displayed Chern number, quantum metric, and quantum volume are evaluated for the single dislocation band $(\ell=u)$ highlighted in Fig.~\ref{Fig2:crystal-dislocations}(a).} In this regime, the redistribution of Berry curvature together with the metric variation drives a reconfiguration of topological band features, enabling dislocations to nucleate chiral magnon modes.  The transition towards trivial states is featured by the jump of Chern number to null values, accompanied by a rapid enhancement of the quantum metric followed by a slow decay. This geometric response thus provides a microscopic link between defect structure, quantum-state geometry, and the emergence of topological magnon channels in nontrivial phases.

\textit{Hexatic phase.--} Having analyzed ordered dislocation arrays, we now consider the disordered limit, generated via Lloyd’s algorithm \cite{Lloyd1982} under periodic boundary conditions. Starting from $N_a$ randomly placed atoms in a domain $L_x \times L_y$, each iteration computes the Voronoi tessellation and moves atoms to the centroids of their cells, including periodic replicas to preserve atom number. After sufficient iterations, the system reaches a stable disordered configuration. Figure \ref{Fig3: Gas-dislocations}(a) and (b) shows dislocations at controlled defect concentrations $\rho$, illustrating low and high densities, respectively, with multiple dislocations of varying lengths and orientations. Crystalline order is monitored via the bond-orientational, ${\cal G}_6(r)=\langle \psi_6(0)\psi_6(r)\rangle$, where $\psi_6({\bs r}_j)=N_j^{-1}\sum_k^{N_j} e^{i6\theta_{jk}}$ is the local 6-fold order parameter with $N_j$ neighbors of $j$ and $\theta_{jk}$ the bond angle to neighbor $k$. At low concentrations, the correlations decay slowly, indicating quasi–long-range order [Fig. \ref{Fig3: Gas-dislocations}(c)]. Increasing dislocation density progressively destroys crystalline order [Fig. \ref{Fig3: Gas-dislocations}(c)]. Beyond a threshold, ${\cal G}_6(r)\sim r^{-\eta}$, showing power-law decay with $\eta=2.5$ and $\eta=3.1$ for concentrations $\rho=0.25$ and $\rho=0.6$, respectively, characteristic of a hexatic phase \cite{Kosterlitz1973,YoungPRB1979,HalperinPRL1978}. This phase emerges when dislocation pairs unbind before disclinations fully destroy orientational coherence during 2D melting \cite{KosterlitzRMP2017}.

Quantum metric has primarily been parametrized by the Bloch momentum \cite{Yu2025}. However, this no longer applies given the lack of translational symmetry, inherent at the presence of randomly distributed dislocations \cite{OliveiraPRB2025}, which is similar to quasicrystals \cite{wang2025,sun2025,Marsal2025,Carrasco2025} and disordered systems \cite{RomeralPRB2025}. The QM is generalized using a real-space approach, analogous to its definition for periodic systems \cite{MarrazzoPRL2019,SousaPRB2023,RomeralPRB2025,Marsal2025,OliveiraPRB2025}. Thus, we substitute the momentum derivatives by commutators $\partial_{k_{\mu}}P\to i[X_{\mu},P]$, and write the corresponding real space expressions for the geometric quantum tensor, $\mathcal{G}=\sum_{\mathbf{r}} g(\mathbf{r})/\mathcal{A}$, where the quantum metric density $g(\mathbf{r})$ is given by 
\begin{equation}\label{eq:G_real_space}
g(\mathbf{r})=-\frac{1}{2\pi} \left([X_{\mu},P][X_{\mu},P]\right)_{\mathbf{r},\mathbf{r}},
\end{equation}
and $\mathcal{A}$ denotes the area of the system. {The projector operator $P$ is associated with the set of eigenstates lying below a given spectral gap of the Hamiltonian, denoted by $\mathcal N$}. It is obtained by using the Bogoliubov transformation T, $P={\text T}\Pi^{(\mathcal{N})}\Sigma_{z}{\text T}^{\dagger}\Sigma_{z}$, where $\Pi^{(\mathcal{N})}$ denotes the matrix with elements $(\Pi^{(n)})_{ij}=\delta_{ij}$ if $i\in \mathcal{N}$ and $(\Pi^{(n)})_{ij}=0$ otherwise. { In Fig.~\ref{Fig3: Gas-dislocations}, the projector $P$ is constructed from all positive-energy
eigenstates below the spectral gap under consideration. For each
dislocation concentration, the spectral gaps are identified from the
real-space spectrum, and the quantum metric is evaluated using the same spectral projector employed in the calculation of the Bott index.} Note that Eq. (\ref{eq:G_real_space}) represents the real-space bosonic analogue of the quantum metric density.

The quantum metric ${\cal G}$, evaluated as a function of dislocation concentration, is shown in Fig. \ref{Fig3: Gas-dislocations}(e) for different pseudo-dipolar strengths $F$. For specific concentrations, $\rho=0.25$ and $\rho=0.6$, the  real-space localization of the quantum metric density, $g(\mathbf{r})$, is shown (highlighted in yellow) at panels (a) and (b), respectively. The behavior of ${\cal G}$ is examined alongside the stability of the topological magnonic phase, quantified by the Bott index ${\cal B}$, a robust real-space invariant of bulk topology \cite{Wang2020,RosalesPRB2024,Loring2010}, displayed at Fig. \ref{Fig3: Gas-dislocations}(f) {for the lower spectral sector}. At low dislocation densities, ${\cal G}$ varies weakly and ${\cal B}$ remains quantized, indicating a stable topological phase. As the dislocation concentration increases, lattice distortions progressively alter the band structure and, beyond a critical threshold $\rho\sim 0.5-0.6$, induce a partial gap closure, see Fig. \ref{Fig3: Gas-dislocations}(d). This transition is accompanied by a sharp enhancement of the quantum metric, signaling the emergence of strongly delocalized and hybridized in-gap modes with significant spectral weight near dislocation cores. Concurrently, the Bott index exhibits a discontinuous jump, marking the transition to topologically trivial states. Beyond this point, at the onset of the hexatic phase, the system becomes dominated by strongly disordered and spatially inhomogeneous states. The concomitant evolution of ${\cal G}$ and ${\cal B}$ establishes a direct correspondence between disorder-induced geometric amplification and the breakdown of band topology.

{\it Conclusions.--} We demonstrate that topological crystalline defects, such as dislocations, can strongly enhance the quantum metric, establishing a direct link between lattice topology and Hilbert-space geometry. This enhancement reshapes magnon band structure and localization, with regions of large quantum metric and altered topological response. We investigate these quantum-geometric effects in both periodic dislocation arrays and in systems lacking translational symmetry. In the latter case, at high defect densities corresponding to the hexatic phase,
magnonic states become topologically trivial and increasingly delocalized, exhibiting strong spatial hybridization. Our results identify crystalline defects as active geometric degrees of freedom, offering a tunable route to control quantum geometry and emergent topology. While the quantum geometric tensor has recently been accessed by spin- and angle-resolved photoemission spectroscopy (ARPES) in electronic systems \cite{Kang2024}, its detection in magnonic platforms remains challenging. Recent progress in nonlinear thermal and charge transport \cite{VarshneyPRB2023,Wang2023}, together with magnetic circular dichroism (MCD) techniques \cite{VinasPRL2023,Koller2025}, provide a viable experimental route for accessing quantum geometry in these systems.
\newpage
{\it Acknowledgements} Funding is acknowledged from Fondecyt Regular 1230747.

\bibliography{QG-bibliography}
\end{document}